# Experimental Up-conversion of Images

Dong-Sheng Ding, Zhi-Yuan Zhou, Wen Huang, Bao-Sen Shi*, Xu-Bo Zou*, and Guang-Can Guo

*Key Laboratory of Quantum Information, University of Science and Technology of China, Hefei 230026,*

*China*

Corresponding author: *drshi@ustc.edu.cn

*xbz@ustc.edu.cn

We experimentally demonstrate the up-conversion of light carrying an image from the infrared spectrum to the visible spectrum using four-wave mixing via a ladder-type configuration in an atomic vapor. The results we obtained show the high correlation between the input image and the up-converted image. We also discuss the possible influences of experimental parameters on the resolution. Our work might be useful for research in astrophysics, night-vision technology and chemical sensing.

## I. INTRODUCTION

Efficient up-conversion of infrared radiation into the visible spectrum has received a great deal of attention recently. This is because infrared CCD cameras often require cooling and suffer from limited spectral response, spatial resolution, or sensitivity. They are also very expensive. Alternatively, up-converted radiation may be detected in the visible spectrum with a highly sensitive low-noise detector with no cryogenic cooling, making up-conversion an attractive technique for detection, imaging, sensing, night-vision and even astrophysical observation [1-4]. For example, the up-conversion technique has been used to observe thermal radiation from the stars for imaging of infrared sources [5], to monitor chemical gas plumes for detecting natural gas leaks and the distribution of greenhouse gases [6], and to transfer the image information to the visible spectrum for optical spectrograms with high resolution [7]. As a result, there is a strong demand for practical schemes that can efficiently convert infrared light into the visible spectrum.

Frequency up-conversion can be realized in nonlinear crystals, but the relatively low conversion efficiency requires a high power laser or a resonant cavity [8-10]. Other promising means are the Raman process [11, 12] and four-wave mixing (FWM) [13] in an atomic ensemble. The up-conversion of a light from the infrared spectrum to the visible spectrum can be realized via a diamond or ladder atomic configuration [14-17]. However, these experiments were achieved without consideration for spatial resolution. So far, there is no experimental report about the up-conversion of an entire image via an atomic system.

In this paper, we discuss the experimental demonstration of the up-conversion of infrared light imprinted as an image to the visible spectrum by FWM in a hot Rubidium vapor. Furthermore, we discuss in detail how the resolution of up-converted images evolves with changes to the experimental parameters. Our results successfully demonstrate image up-conversion via an atomic system. Our setup is simple compared to the setups performed with a nonlinear crystal: neither a resonant cavity nor a high power laser is required [8-10], and also simple compared with the scheme of an up-conversion (without an image) performed in a cold atomic ensemble [15], where laser trapping and a high vacuum are needed. Another big advantage is that our experimental setup could be reduced in size and is mobile.

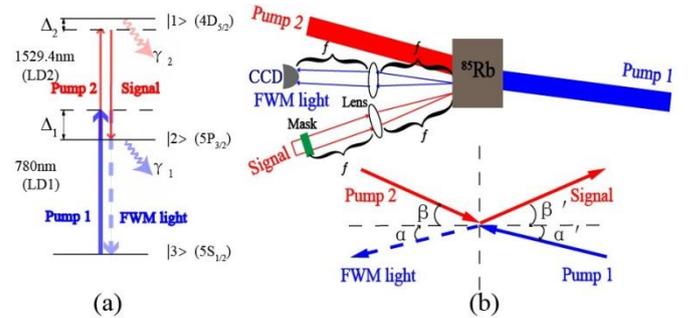

Fig. 1. (Color online) (a): Energy level diagram of the ladder-type configuration used in the experiment. (b): Upper is a schematic diagram of the experimental set-up for up-conversion process. Bottom is the phase matching condition, $\beta \approx 2\alpha$, $\alpha=\alpha'$, $\beta=\beta'$. The red line indicates the paths of the 1529.4 nm light, and the blue line that of the 780 nm light.

## II EXPERIMENTAL CONFIGURATION AND THEORETICAL ANALYSIS

A ladder-type configuration of $^{85}$Rb atom used in our experiment is shown in Fig. 1(a). It consists of one ground state |3>, one intermediate state |2> and one upper state |1>. They are the $5S_{1/2}$, $5P_{3/2}$ and $4D_{5/2}$ levels of $^{85}$Rb, respectively. The transition frequency between $5S_{1/2}$ and $5P_{3/2}$ corresponds to the $D_2$ line (780 nm) of $Rb^{85}$, and the transition between $5P_{3/2}$ and $4D_{5/2}$ can be coupled by a laser at 1529.4 nm. $\Delta_1$ (The blue-shifted detuning represents positive.) is the detuning of the transition $5S_{1/2}(F=3)$-$5P_{3/2}(F'=3)$, and $\Delta_2$ (The red-shifted detuning represents positive.) is the detuning of the transition $5P_{3/2}(F'=3)$-$4D_{5/2}(F''=4)$. $\gamma_{1(2)}$ is the decay rate of the energy level |2> (|1>). A simplified experimental setup is illustrated in Fig. 1(b). A 5-cm $^{85}$Rb vapor cell containing isotropically-pure $^{85}$Rb is used as an up-converter. The diameter of the cell is 2.5 cm. A 780 nm CW laser polarized in the horizontal direction, from an external-cavity diode laser (LD1, DL100, Toptica), is input to the Rb cell as pump 1. A 1529.4 nm CW laser from another external-cavity diode laser (LD2, DL100, Prodesign, Toptica) is divided into pump 2 and the signal by a beam splitter. These two beams nearly co-propagate and have horizontal and vertical polarization directions, respectively. The angle between these two beams is 2.15°. Pump 1 nearly counter-propagates with pump 2. The small angle between these two beams is 0.86°. The signal beam passes through an image mask. The up-converted field at 780 nm with the same polarization as pump 1 nearly counter-propagates along the signal direction, and is monitored by a common CCD camera. A 4-$F$ imaging system is used to project the object of the mask onto the camera. It consists of two lenses, each of focal length $F$=500 mm separated by a distance $2F$. The mask is placed at the front focus of the first lens and the image is obtained at the back focal plane of the second lens. The vapor cell is placed at the back focal plane of the first lens. The waist of the signal beam at the center of the cell is 0.78 mm. Pumps 1 and 2 are weakly focused and their diameters at the center of the cell are 5.2 mm and 4.6 mm, respectively. All beams are spatially filtered by a single mode fiber, therefore they are monomode Gaussian TEM00. This non-collinear FWM configuration is used in the experiment so that the up-converted image is easily separated from the pump fields, therefore no spectral filtering is necessary.

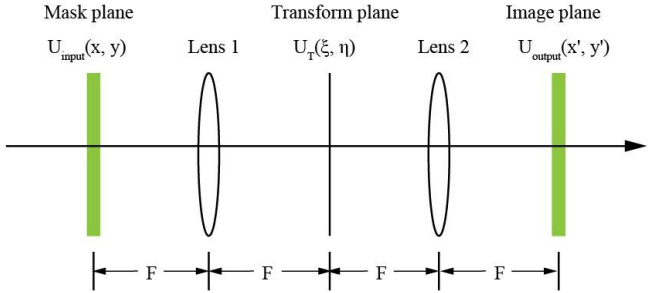

Fig. 2. (Color online) 4-F system configuration for conversion of infrared image

Before introducing our experimental results, we give a simple theoretical description of our system. The simplified diagram is shown in Fig. 2. The 4-$F$ system consists of a mask plane, a transform plane, an image plane, and two lenses (lens 1 and lens 2). $U_{input}(x,y)$ stands for the image at mask plane, $U_T(\xi,\eta)$ is at the transform plane and $U_{output}(x',y')$ is at the image plane. $x, y, \xi, \eta, x'$ and $y'$ represent two dimensional coordinates.

Through lens 1 with a focal length of F, the diffracted image at the transform plane can be expressed as:

$$U_T(\xi,\eta) \propto \iint U_{input}(x,y)\exp(-i\bar{k}_S \frac{x\xi+y\eta}{F})dxdy \quad (1)$$

Considering the phase-matching of the FWM processes, we obtain the generated diffracted image at the transform plane as below:

$$U'_T(\xi,\eta) \propto \iint U_{input}(x,y)\exp(-i\bar{k}_F \frac{x\xi+y\eta}{F})dxdy \quad (2)$$

where, the wave numbers satisfy $\bar{k}_F = \bar{k}_2 + \bar{k}_S - \bar{k}_1$ due to the conservation of momentum of the FWM process, $\bar{k}_{1,2,S,F}$ are the wave vectors of pump 1, pump 2, the signal and the up-converted fields. The inverse Fourier transform of Eq. (2) is:

$$U_{input}(x,y) \propto \iint U'_T(\xi,\eta)\exp(i\bar{k}_F \frac{x\xi+y\eta}{F})d\xi d\eta \quad (3)$$

After passing through lens 2, with focus length of F, the image at the image plane can be expressed as:

$$U_{output}(x',y') \propto \iint U'_T(\xi,\eta)\exp(-i\bar{k}_F \frac{x'\xi+y'\eta}{F})d\xi d\eta \quad (4)$$

We can obtain the equation: $U_{output}(x',y') \propto U_{input}(-x,-y)$. This illustrates the output is a conjugate image to the input image.

## III. RESULTS

As a principle proof, we perform the up-conversion experiment with a mask imprinted with the digits "0", "2", "3", "4", "5" and "6". Fig. 3 is a set of experimental data under the following conditions: powers of the signal, pump 1 and pump 2 are 0.15, 2 and 9.3 mW respectively; the cell is heated to 139 ℃, providing a rubidium vapor density of $5.6\times10^{13}$ cm$^{-3}$. The detuning $\Delta_1$ and $\Delta_2$ of the 780 nm and 1530 nm lasers is 1.5 GHz and 794 MHz, respectively. The reason why we take the large single-photon detuning both at pump 1 and pump 2 (and the signal) in our experiment is to significantly reduce the Rb cell's resonant absorptions of the up-converted image and signal. In Fig. 3, (a) is the direct image of the mask illuminated by the signal beam without the Rb cell, (b) is the image imprinted at signal beam obtained after it passes through the Rb cell and (c) is the up-converted image. As stated by Eq. (4), Fig. 3 shows the successful up-conversion of an image. The main features of the image have been preserved during this nonlinear FWM process. The up-converted images are conjugate relative to input images.

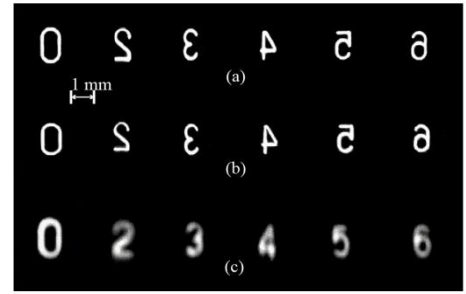

Fig. 3. A set of experimental data. (a): the direct image of the mask imprinted at signal beam obtained without the Rb cell; (b): the image imprinted at signal beam obtained after it passes through the Rb cell; (c): the up-converted image.

In Fig. 3(c), the up-converted images are unclear, and the edges are softened in comparison with Fig. 3(b). There are several limits which affect the resolution of the up-converted image: one limit is the diffractive effect of light during the propagation. The image imprinted by the signal can be decomposed into a set of plane waves; each plane wave acquires different phases in the FWM process. The superposition of each generated plane wave composes the FWM image, which becomes bold and blurry due to the difference in phase. Further, the Rb media acts as a spatial filter due to phase matching in the FWM process, which obstructs the up-conversion of high frequencies.

In the following, we take the digit "2" as an example to check the influences of some experimental parameters on the resolution of the up-converted image. First, we check the influences from the single-photon detuning of the pump lasers. In the experiment, we change the detuning of the 780 nm laser and 1530 nm laser to see how the up-converted image evolves, while keeping the other parameters unchanged. The experimental results are shown in (a) and (c) of Fig. 4, where the evolution of the up-converted images with single-photon detuning is shown. The experimental results show us that single-photon detuning has no obvious effect on the resolution of an up-converted image; it only affects the efficiency of the up-conversion. However, this is true only if the power of the pump is small. If it is large, an image phase mismatch will occur, and a different phenomenon will appear, which is discussed below.

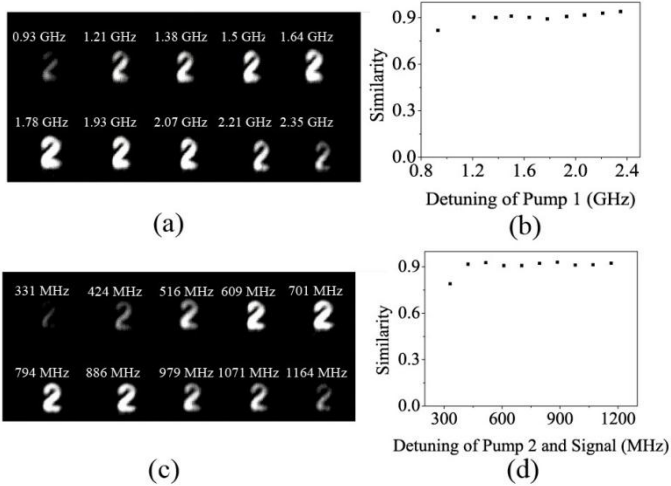

Fig. 4. (a) and (c) show how the up-converted images evolve with the detuning of the 780 nm and 1530 nm laser, respectively under these conditions: Powers of the signal, pump 1, and pump 2 are 0.15, 2 and 9.3 mW, respectively. The detuning of the 1530 nm laser is 794 MHz in (a) and of the 780 nm laser is 1.5 GHz. The temperature of the cell is 139 ℃. (b) and (d) give the calculated similarity between the original image and the up-converted image.

In order to give a more accurate and quantitative evaluation of the transfer process, we calculate the similarity $R$ between the original image and the up-converted image by the equation:

$$R = \frac{\sum_m \sum_n (A_{mn})(B_{mn})}{\sqrt{(\sum_m \sum_n (A_{mn})^2)(\sum_m \sum_n (B_{mn})^2)}}, \quad (5)$$

which is referenced by Ref. [18]. Where, $A_{mn}$ is the intensity recorded for pixel (m, n) of the input image and $B_{mn}$ is the intensity recorded for pixel (m, n) of the up-converted image. Figs. 4 (b) and (d) show the calculated similarity between the original image and the up-converted image. We could find that the similarity is at the level of 0.9 in most cases and is almost independent of the single-photon detuning of the pump. The dip in similarity in Figs. 4(b) and (d) at the small single-photon detuning end is due to the inefficiency of up-conversion. In order to give qualitative statements about the similarities, we compare the value of the similarity of number "2" and number "3" and "4" by the factor R=0.78 and R=0.64.

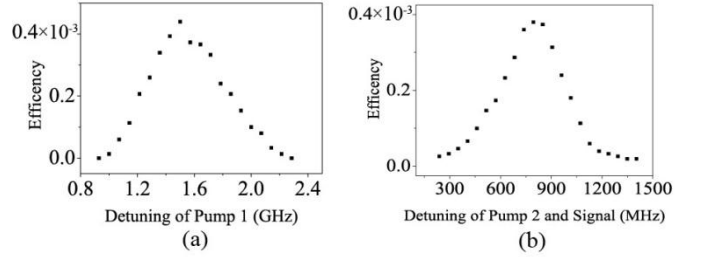

Fig. 5. (a) The conversion efficiency against the single-photon detuning of a 780 nm laser under the condition of $\Delta_1$ =794 MHz and (b) The conversion efficiency against the single-photon detuning of a 1530 nm laser when $\Delta_2$=1.5 GHz. The other parameters are the same as in Fig. 3.

We also measure how the conversion efficiency evolves with single-photon detuning, and the experimental data is shown in Fig. 5. The data clearly shows that the single-photon detuning strongly influences the up-conversion efficiency. There is a trade-off between single-photon detuning and the efficiency: if the detuning is too large, then the FWM becomes too weak, and the efficiency of up-conversion is small. If the detuning is too small, then the large absorption rate the Rb has on the up-converted image and the infrared signal also decrease the efficiency of up-conversion. Near two-photon resonance, the increased energy density in slow light also can increase the efficiency of up-conversion. The maximal conversion efficiency obtained is $4.5 \times 10^{-4}$ and is much lower than the 50% obtained with the strong pump fields in our Lab (see Ref. 19, where the angle between the pumps are small, and the power density of pumps are large, so the nonlinear interaction is strong). We conclude that the power density of the pump should not be too large, otherwise the self-focusing and self-defocusing [20] effects will appear, which greatly affect the quality of up-converted image. Another possible limit on the efficiency of up-conversion is the geometry of the setup: the non-collinear configuration may prevent appreciable coupling among the four beams over the length of the cell. A qualitative theoretical analysis about the up-conversion efficiency in a diamond-type configuration is given in Ref. 14, and a quantitative calculation for a lambda-configuration can be found in Ref. 21. In Ref. 19, there is a quantitative analysis of the up-conversion efficiency.

## IV. PATTERN FORMATION

If we change the power of pump 1, we find that an undesired pattern formation appears when the power is too large. Fig. 6 is the experimental data obtained. Such a phenomenon strongly affects the quality of the up-converted image, and therefore should be avoided for any use of up-conversion. A similar phenomenon has also been observed and discussed in Ref. [22, 23, 24] in detail, in which a spatial pattern is formed. The results show that the appearance of these effects is strongly dependent on the power density of the laser. If the phase matching condition is satisfied, the input image is converted into the output image with high resolution. The pattern appears when the image phase matching condition $\bar{k}_F = \bar{k}_2 + \bar{k}_s - \bar{k}_1$ is not satisfied as $\bar{k}_1$ is distorted by a spatial index of refraction in the atom vapor cell. It is easy to compensate by changing some experimental conditions, for example, by changing the focus condition of pump 1 (from parallel to ~1000 mm focus length) or the focus condition of the pump 2. The compensated image is the last image listed in Fig. 6.

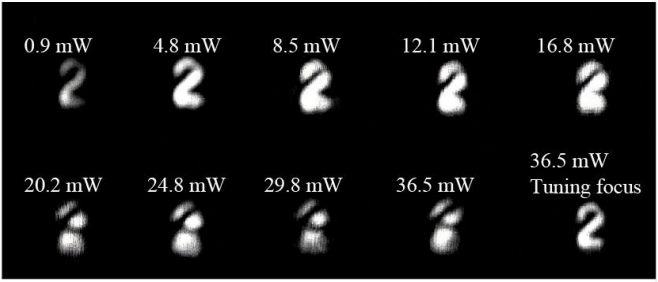

Fig. 6. The observed up-converted image evolves as the power of pump 1 increases. The other experimental parameters are held constant.

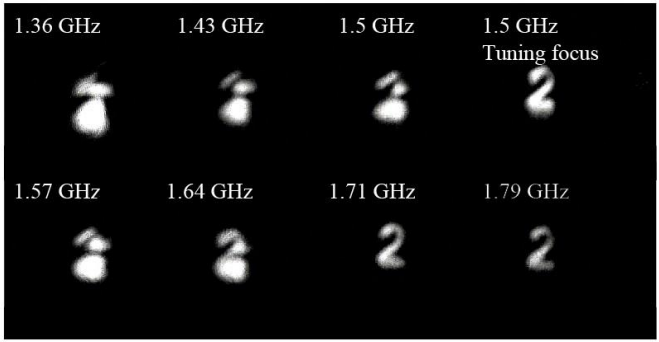

Fig. 7. The observed up-converted image evolves as the detuning of pump 1 decreases. In this experiment, the power of pump 1 is 30 mW. All other parameters are held constant.

A similar spatial phase mismatching phenomenon is also observed when we change the single-photon detuning of the 780 nm laser, if the power of pump 1 is quite large. The results are shown in Fig. 7. We find that the up-converted image breaks into a pattern with decreased detuning. The vector $\bar{k}_1$ is altered due to the different frequencies of propagation, which results in a phase mismatch. By a method similar to that mentioned above, we can also compensate for this spatial phase mismatch by changing some parameters, such as adjusting the focusing conditions of the laser. (The compensated image is the last image in the top row of Fig. 7)

## V. DISCUSSION

We want to point out that this process of up-converting an image is linear under suitable conditions. We have experimentally demonstrated this distinguishing feature by up-converting orbit angular momentum [25]. Another work is Ref. [26] which report the transfer of phase structure using orbital angular momentum. We want to mention that although Lett *et. al.* reported on the generation of spatially multimode twin beams using FWM in a hot atomic vapor [27, 28] and Tabosa *et. al.* [29] reported on an experiment about multimode transfer through a FWM process, their experiment showed significant differences from ours: first, a lambda-type atomic structure is used in their FWM experiment; second, all four beams in their FWM are nearly of the same frequency. We want to emphasize here that although our experiment is done with a special ladder-type configuration, the up-conversion of an image from 1530 nm to 780 nm is realized, and the demonstrated technique can be further adapted for other ladder-type configurations of the Rb atom or for the different atomic systems. Therefore this technique can be used to up-convert an image for a wider infrared wavelength range.

## VI. SUMMARY

In summary, we demonstrate experimentally the up-conversion of images from infrared radiation to the visible spectrum by using FWM in a hot Rb vapor. The possible influences on the resolution of the up-converted image are discussed. Our research results are useful in research in optics image processing.

**Acknowledgments**

This work was supported by the National Natural Science Foundation of China (Grant Nos. 10874171, 11174271), the National Fundamental Research Program of China (Grant No. 2011CB00200), the Innovation fund from CAS, Program for NCET.